# Temporal Shifts and Relationships of Emotions in Social and Mass Media: A Case Study of the "Reiwa Rice Riot" in Japan

Erina Murata [a,b*], Masaki Chujyo [a] and Fujio Toriumi [a*]

a School of Engineering, The University of Tokyo, 7-3-1 Hongo, Bunkyo-ku, Tokyo, 113-8656, Japan. (e-mail: erinamurata@g.ecc.u-tokyo.ac.jp).

b School of Human Sciences, Waseda University, 2-579-15 Mikajima, Tokorozawa, Saitama, 359-1192, Japan. (e-mail: erinamurata@ruri.waseda.jp).

* Corresponding author: Fujio Toriumi, School of Engineering, The University of Tokyo, Japan. (e-mail: tori@sys.t.u-tokyo.ac.jp).

In Japan, severe rice shortages in 2024 sparked widespread public controversy across both news media and social platforms, culminating in what has been termed the "Reiwa Rice Riot." This study proposes a framework to analyze the temporal dynamics and directional relationships of emotions expressed on X (formerly Twitter) and in news articles, using the "Reiwa Rice Riot" as a case study. While recent studies have shown that emotions are dynamically associated across social and mass media, the patterns and pathways of such emotional shifts remain insufficiently understood. To address this gap, we applied a machine learning–based emotion classification grounded in Plutchik's eight basic emotions to analyze posts from X and domestic news articles. Our findings suggest that emotional shifts on X tended to precede those in news media in a temporal sense. Furthermore, in both media platforms, fear was initially the most dominant emotion, but over time intersected with anticipation which ultimately became the prevailing emotion. Our findings suggest that patterns in emotional expressions on social media may serve as a lens for understanding temporal shifts in emotions during social crises and the temporal precedence of emotions across social and news media.

## Introduction

In recent years, the widespread use of social media has enabled individuals to transmit and share information online.[1] As a result, communication surrounding specific social phenomena and events has come to exhibit dynamics that differ from those observed in traditional mass media-driven communication.[2] Traditionally, the propagation of information has been led by mass media, for example by shaping public awareness of a topic's importance by repeatedly reporting on it.[3]

However, in recent years, social media have also come to play an important role in information diffusion and online discussions. Among these platforms, X (formerly Twitter) has attracted particular attention as a space where diverse opinions and emotions are exchanged in real time.[4] As such, it serves as a valuable data source for capturing visible emotional expressions and discussions related to socially important issues. Clarifying the temporal relationships between mass media and X is therefore important for understanding how information and emotional expressions evolve across media environments.

We focus specifically on emotional dynamics rather than issue salience, as emotions are rapidly disseminated and amplified across media environments.[5] For instance, the continuous dissemination of news concerning COVID-19 infection and mortality rates has been suggested to amplify the emotion of social anxiety and potentially exert detrimental effects on mental health. [6] In addition, emotions influence information exchange across media, particularly through the bidirectional interactions between social and mass media.[7-8] For example, during the COVID-19 pandemic, social media functioned as an important communication tool that made citizens' voices visible.[9] Emotions such as anxiety and anger regarding the infection situation and government responses were shared on social media,[10] and as these sentiments gained attention, mass media followed suit by focusing on similar emotional themes. Conversely, repeated circulation of news content on social media further intensified negative emotions.[11] These processes can be understood as mechanisms of emotional expression, amplification, and feedback across media environments.

Despite these observations, it remains unclear whether emotional expressions exhibit systematic intermedia dynamics. Social media and mass media are often observed to co-evolve in the dissemination of emotions, and understanding how emotions emerge and evolve across platforms is crucial for understanding emotional dynamics in digital communication environments. However, systematic methods for quantifying and visualizing intermedia emotional dynamics—particularly those based on multi-emotion models—remain underdeveloped. Additionally, prior studies have explored how information and emotions diffuse during disasters, elections, and social movements,[12–15] but few have systematically examined how emotions themselves evolve over time in large-scale, rapidly unfolding events. Insights into the rapid spread of emotions such as anxiety and fear can support early anomaly detection and guide timely, well-toned information strategies[16]. Accordingly, this study explores whether directional relationships can be observed in emotional time series across media.

This research conducts an analysis using the "Reiwa Rice Riot" as a case study. The "Reiwa Rice Riot" refers to a public controversy in Japan in 2024 triggered by a shortage in rice supply and a surge in rice prices. The incident was caused by a confluence of multiple factors, including a reduction in rice harvests due to abnormal weather conditions and increased demand for rice following wheat price hikes stemming from the war in Ukraine. A timeline of major developments during the "Reiwa Rice Riot" is provided in Appendix A[17-20]. In Japan, white rice is a traditional staple food, and historically, rice shortages have acted as triggers for social unrest, owing to cultural and historical contexts. During the "Reiwa Rice Riot", the lack of foreseeable recovery in rice supply led to public anxiety and confusion. The incident had tangible impacts on market trends and government responses,[21-22] and it is regarded as a typical example in which emotion functioned as a trigger for information dissemination. Furthermore, the incident was widely covered across both social media and news media, with information spreading concurrently on both platforms. For this reason, this study positions the "Reiwa Rice Riot" as a public controversy marked by significant emotional volatility and examines it as a valuable case for analyzing directional emotional relationships between social and news media. To analyze this phenomenon, we adopt Plutchik's eight basic emotions [23] as the analytical framework for emotion classification. Plutchik's eight basic emotions is rooted in an evolutionary psychological perspective and is widely used in corpus construction and text-based emotion analysis.[24-26]

We develop a framework to elucidate the temporal changes of emotions and the directional relationships between them in social media and news media. Specifically, we applied a machine learning–based emotion classification model grounded in Plutchik's eight basic emotions to Japanese-language texts related to the "Reiwa Rice Riot" collected from both social media and news sources. Using the estimated emotion scores, we conducted time-series analyses to examine emotional trends and relationships between media. The results revealed a clear shift in both social media and news media from predominantly negative emotions such as fear to more positive emotions such as anticipation. Notably, this shift occurred earlier on social media than in news media, a precedence further corroborated through transfer entropy analysis. These findings suggest that social media may exhibit earlier emotional changes than news media in certain cases. The primary contribution of this study lies in presenting a systematic method for analyzing intermedia emotional relationships and in empirically demonstrating the precedence of social media in our case study.

## Methods

This section first explains the acquisition of social and mass media data and subsequently introduces the proposed analytical framework of this study. An overview of the framework is provided in Figure 1.

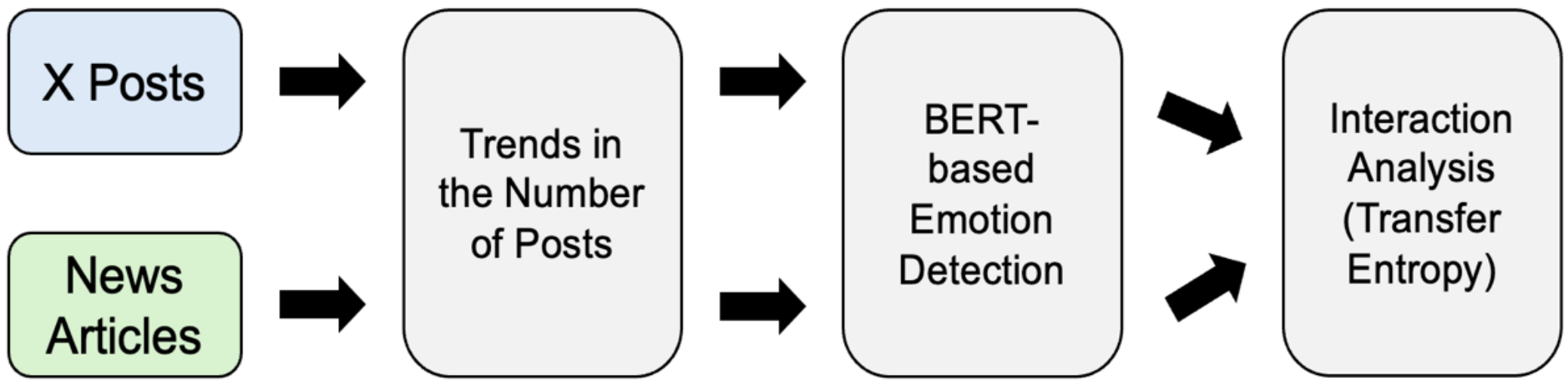

**Figure 1.** The proposed analytical framework of this study

### Data Collection from X

We collected X posts containing the keywords “rice shortage” or “shortage of rice.” Retweets were excluded to include only original posts, and non-Japanese language tweets were also excluded. The data collection period was two months, from August 1, 2024, at 00:00 (UTC+9) to September 30, 2024, at 23:59 (UTC+9). The period from August to September 2024 was selected to cover the main phase of public attention and controversy surrounding the “Reiwa Rice Riot.” Using the API of X, a total of 156,000 posts were collected and used as the primary data for analysis.

## Data Collection from News Articles

Using Ceek.jp News,[27] we collected news articles that contained either the keyword "rice shortage" or "shortage of rice." Ceek.jp News is a news search engine that crawls news articles from domestic Japanese media sources. The collection period was also from August 1 to September 30, 2024, matching the X data period. A total of 1,034 articles were retrieved, and both titles and body texts were included in the analysis.

## Framework of This Study

We adopt an integrated analytical framework to examine the temporal shifts and relationships of emotions across social and mass media. The framework is grounded in Plutchik's psych evolutionary theory of eight basic emotions, which allows for a multidimensional view of affective processes beyond simple polarity. Emotions embedded in textual data are estimated with a machine learning–based classifier and then processed as continuous time series, enabling the detection of shifts in dominant emotions over time. To investigate directional relationships between media platforms, transfer entropy analysis is employed, providing a systematic means to examine the temporal precedence and directional patterns of emotional change.

## Emotion detection by BERT-based model

To analyze the temporal evolution of emotions, we constructed a BERT model using the WRIME dataset. The WRIME dataset, developed by Kajiwara et al.,[28] is a Japanese-language dataset designed for research in emotion analysis. In WRIME, each tweet is annotated with Plutchik's eight basic emotions (fear, sadness, surprise, anticipation, joy, anger, disgust, trust), and each emotion is further labeled with four levels of intensity: none, weak, moderate, and strong.

We fine-tuned a pre-trained Japanese BERT model[29] by adding a linear layer on top of the 768-dimensional CLS token representations, using normalized intensity vectors in WRIME as target labels. The main training settings were as follows: a batch size of 8 (per_device_train_batch_size = 8), 1 training epoch (num_train_epochs = 1.0), evaluation at every 200 steps (evaluation_strategy = "steps", eval_steps = 200). The learning rate was set to the default value of the Transformers library. Input texts were tokenized with truncation and padding to a fixed length (truncation = True, padding = "max_length"). We did not perform extensive hyperparameter tuning; instead, the model was trained primarily using standard settings. A Softmax function was applied to output the probability distribution for each of the eight emotions. The model performance on X was evaluated, achieving approximately 75% accuracy in predicting the dominant emotion. (see Appendix B) In addition, the model was further evaluated on news articles by comparing the dominant emotions identified by human annotations and model predictions, with an agreement rate of 53% (see Appendix C).

For the time series analysis, we first computed the daily mean score for each emotion across all posts and news articles. Subsequently, the 7-day moving average was obtained by averaging these daily scores over consecutive 7-day windows, with the window advanced by one day across the observation period.

### Directional Relationship Analysis Using Transfer Entropy

To examine whether there were directional relationships between emotional time series extracted from X and news articles, we calculated transfer entropy.[30] Prior to the transfer entropy analysis, stationarity of the time series was examined using the Augmented Dickey–Fuller (ADF)[31], KPSS tests[32]. Transfer entropy is a metric that quantifies the extent to which the past states of one time series contribute to the prediction of the future states of another series. In order to estimate probability distributions, emotion scores were discretized using equal-width binning into five bins based on the minimum and maximum values of each time series. This approach preserves ordinal relationships in the data while enabling discrete-state estimation of transfer entropy. Subsequently, bidirectional transfer entropy was computed using the discretized labels for both X and news data. In this study, we constructed the time series using only the dates for which both news and social media data were available, excluding dates with missing values. Equation (1) represents the transfer entropy from X to news:

$$TE_{X\to Y} = \sum_{y_t,\, y_{t-1},\, x_{t-1}} P(y_t, y_{t-1}, x_{t-1}) \log \frac{P(y_t \mid y_{t-1}, x_{t-1})}{P(y_t \mid y_{t-1})} \tag{1}$$

where $x_t$ and $x_{t-1}$ denote the discretized emotion labels from X at time t and t-1, respectively, and $y_t$ and $y_{t-1}$ represent the corresponding labels for news articles.

$P(y_t, y_{t-1}, x_{t-1})$ represents the empirical probability distribution derived from the frequency of discretized categorical values.

$P(y_t \mid y_{t-1}, x_{t-1})$ denotes the conditional probability of observing emotion label $y_t$ at time t in news articles given that the label was $y_{t-1}$ at the previous time step and the corresponding X label was $x_{t-1}$.

Similarly, $P(y_t \mid y_{t-1})$ represents the conditional probability of $y_t$ given only the preceding news emotion label $y_{t-1}$.

By calculating Equation (1), we quantify the directional dependence between X and news. A similar procedure was applied to calculate the transfer entropy from news to X. To explore temporal variations in emotional transfer entropy, we conducted additional analyses. Specifically, we first calculated transfer entropy separately for August and September. Furthermore, we applied a sliding-window transfer entropy analysis using a 14-day moving window with a one-day step to capture time-varying relationships.

## Results

### Trends in the Number of Posts

Figure 2 shows the 7-day moving averages of the number of posts on X and the number of news articles. The moving average is employed to smooth fluctuations arising from day-of-week effects and to provide a clearer depiction of the overall trends in information dissemination and convergence.

Posts on X are indicated in blue, and news articles are shown in green. The peak in the number of posts on X occurred several days earlier than the peak in news article counts. The number of posts on X reached its peak in late August 2024, while news articles peaked in early September. Subsequently, both media showed a calming trend toward the end of September, with the moving averages stabilizing.

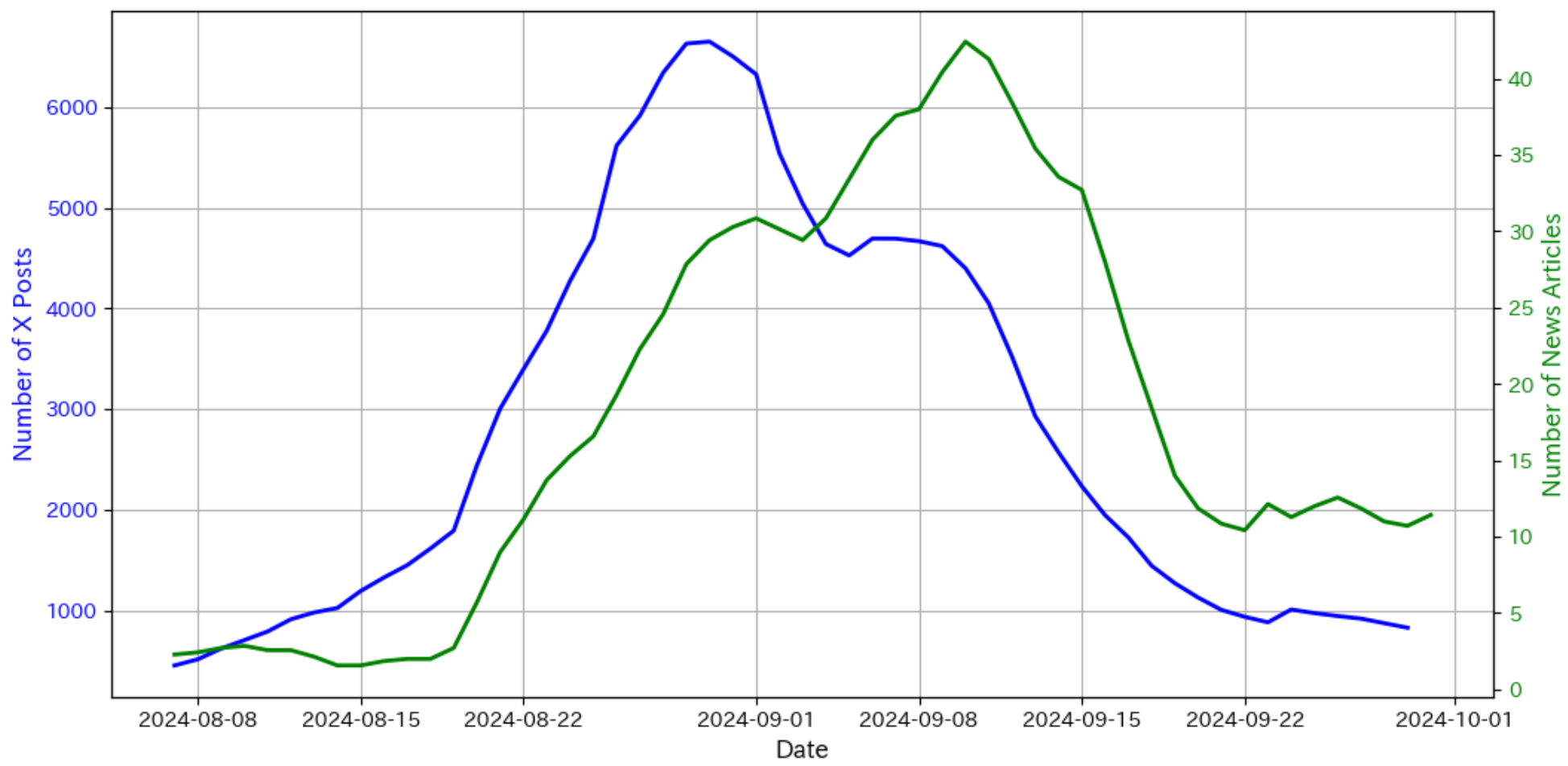

**Figure 2.** 7-day moving average of the number of posts on X and the number of news articles

### Emotion detection by BERT-based model

Scores for each of Plutchik's eight basic emotions were calculated, and the 7-day moving average time-series plots were created. The emotional trends on X are shown in Figure 3, and those in news articles are presented in Figure 4. [1] The colors represent the following emotions: fear (blue), sadness (orange), surprise (green), anticipation (red), joy (purple), anger (brown), disgust (pink) and trust (gray).

On X, fear had the highest score in the initial phase and peaked in mid-August. In contrast, the score for anticipation gradually increased, surpassing fear in early September to become the most dominant emotion. This visualized a shift from initially dominant negative emotions such as fear and sadness to increasingly stronger positive emotions such as anticipation and joy.

Furthermore, in both X and news media, fear was initially the strongest emotion, which later intersected with anticipation, with anticipation eventually becoming the most prominent. We also find that the day on which the moving averages of fear and anticipation crossed occurred earlier on X than in news media.

[1] Note that the emotion classification has an accuracy of 75%, which may introduce noise particularly for less frequent emotions. While dominant emotional trends are expected to be robust, the interpretation of minor emotional components should be treated with caution.

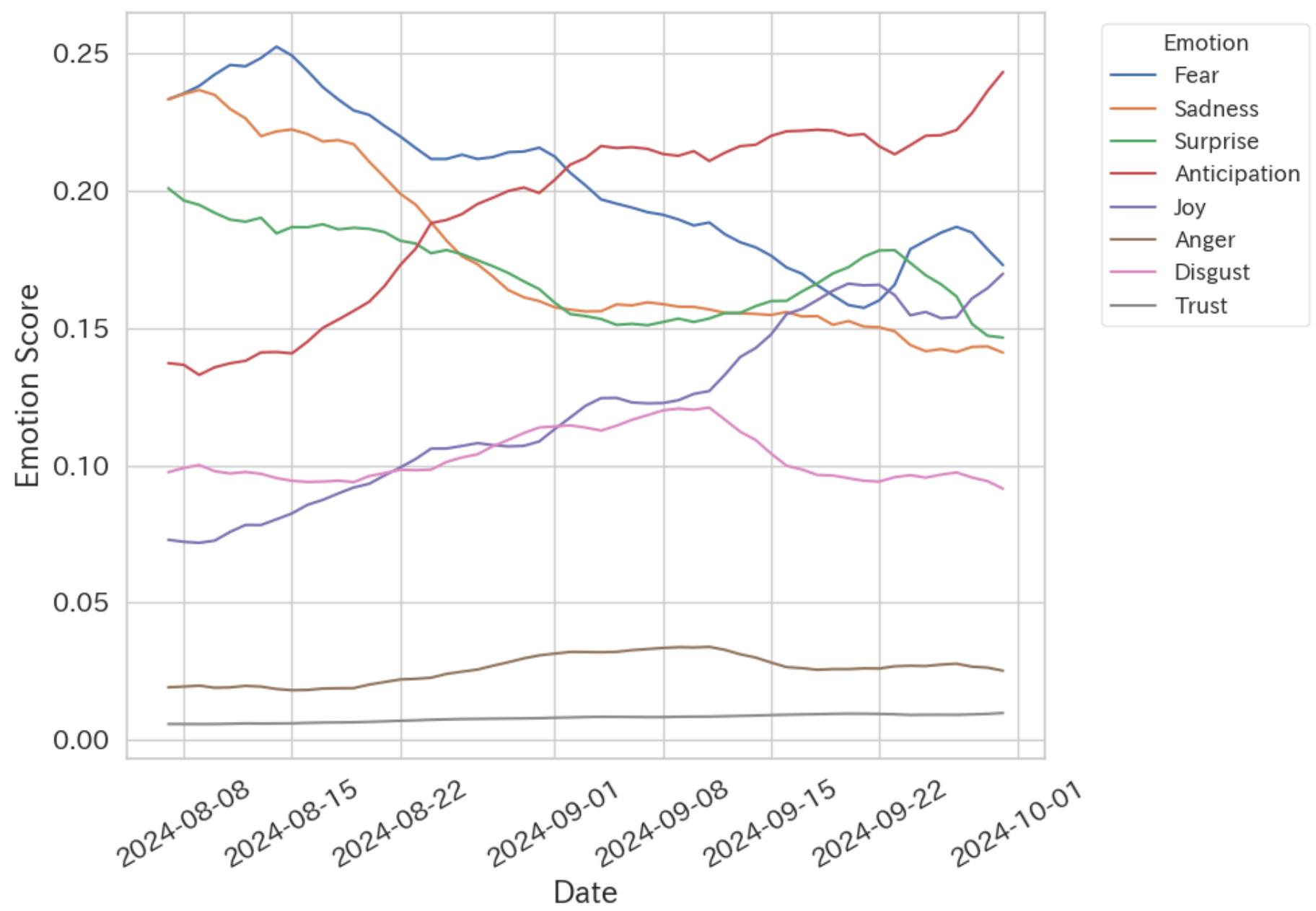

**Figure 3.** Emotion scores on X

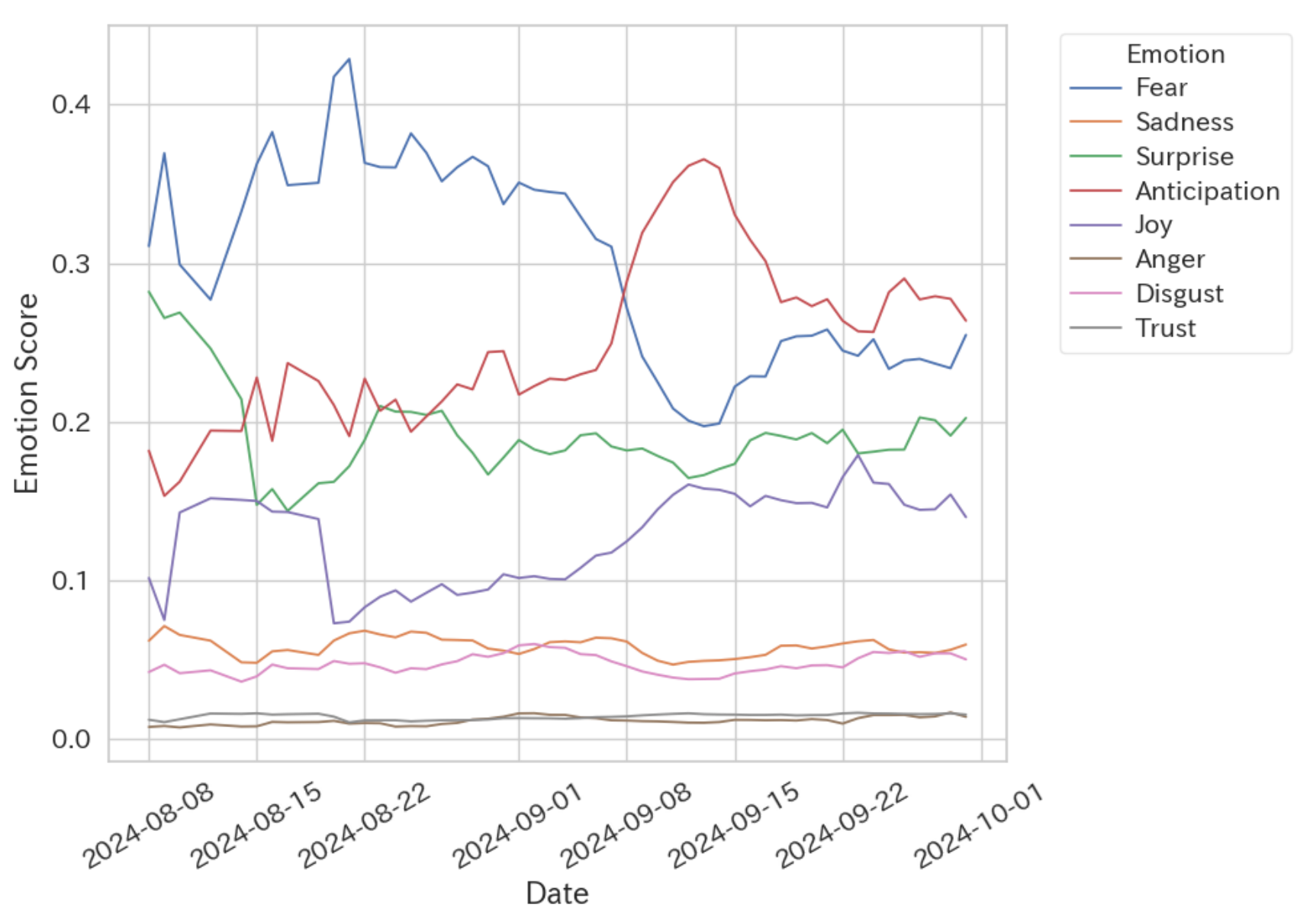

**Figure 4.** Emotion scores in news articles

**Directional Relationship Analysis Using Transfer Entropy**

Figures 2-4 suggest that both the peak in the number of posts and in emotional intensity occurred earlier on X than in the news. To quantitatively assess the directional relationships in emotional trends, we calculated the transfer entropy for each emotional category between X and news articles. Prior to this analysis, stationarity tests using the ADF and KPSS tests revealed that most news-related emotional time series were stationary, whereas X-based series were largely non-stationary. Given these properties of the data, we first applied transfer entropy to the full time series and, due to the presence of non-stationarity, further conducted a sliding-window transfer entropy analysis to account for time-varying relationships.

Figure 5 shows the transfer entropy for each emotion. Blue bars indicate transfer from news to X, and orange bars represent transfer from X to news. In all emotion categories, the values for from X to news exceeded those for from news to X. Particularly high transfer entropy was observed for anger, disgust, and fear. This suggests that increases in negative emotions on X tended to temporally precede similar emotional patterns observed in news articles. As a robustness check, we varied the number of bins used for discretization and confirmed that the overall directional patterns remained largely consistent across different settings; detailed results are provided in Appendix D.

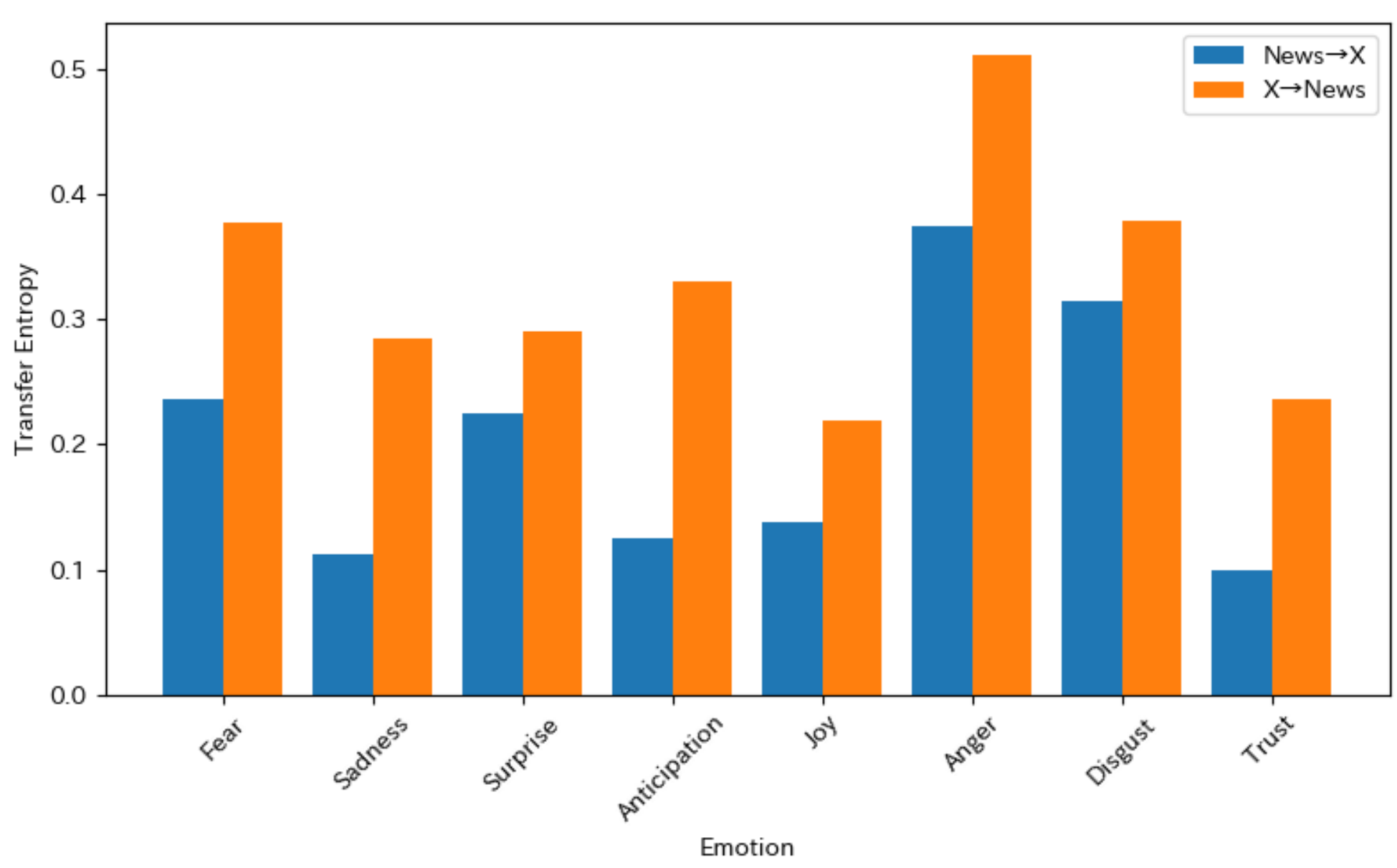

**Figure 5.** Transfer entropy by emotion category

To further examine how emotional transfer entropy changed over time, we conducted separate calculations for August and September. The results for August are shown in Figure 6, and those for September in Figure 7.

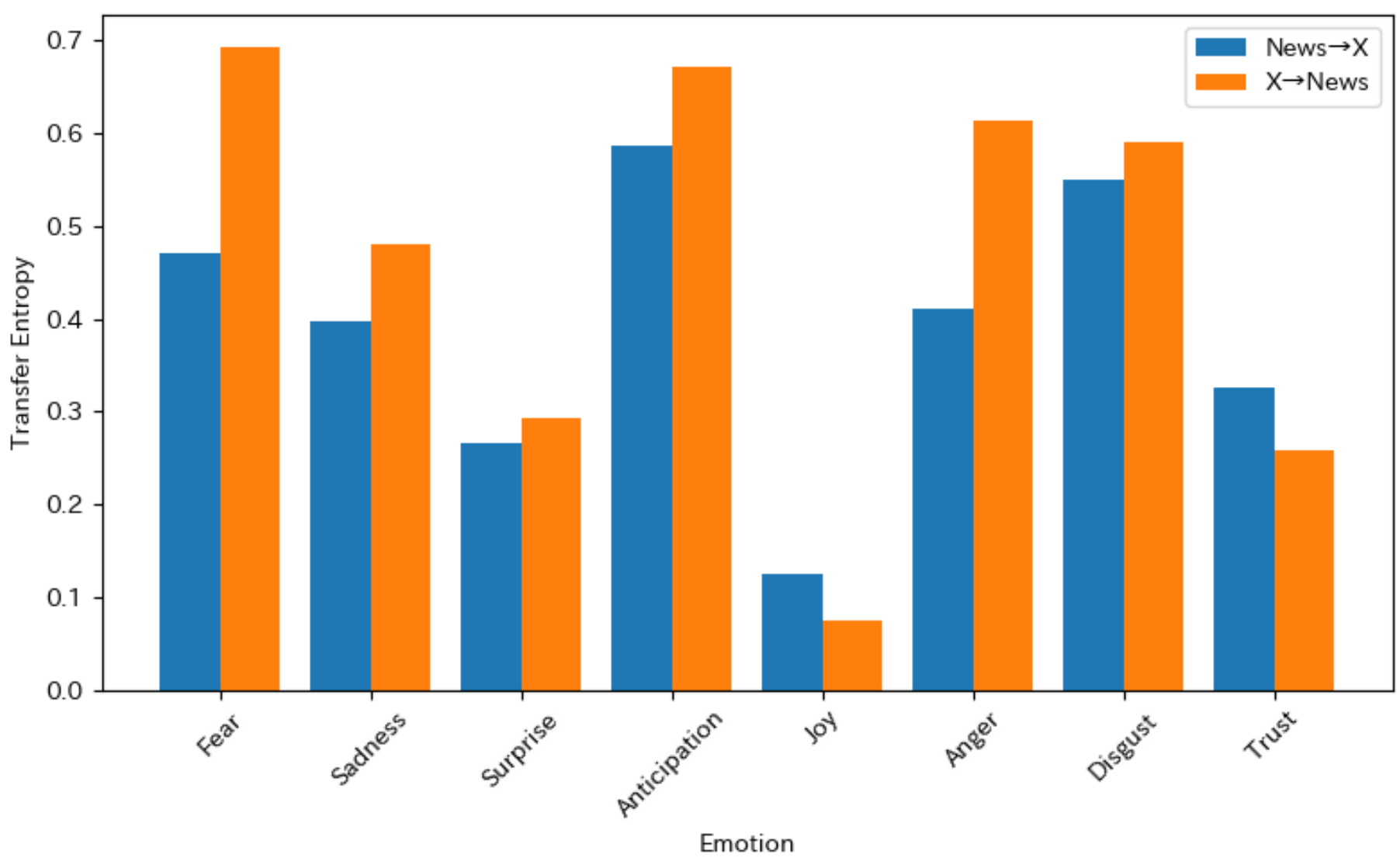

**Figure 6.** Transfer entropy by emotion category (August)

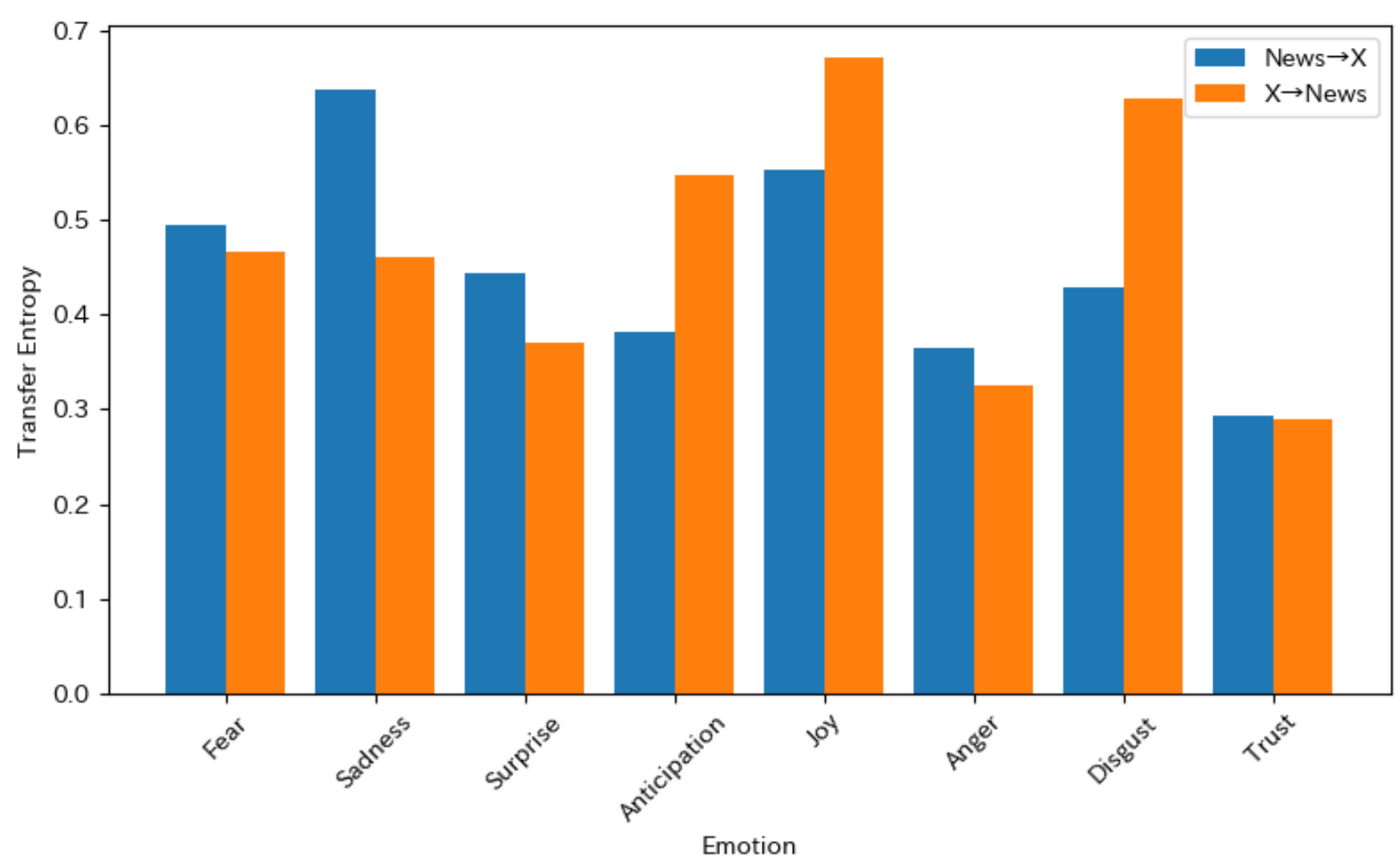

**Figure 7.** Transfer entropy by emotion category (September)

In August, for most emotions, the transfer entropy from X to news was greater. However, for joy and trust, the transfer from news to X was larger. Notably, in August, anger and fear exhibited high transfer entropy from X to news, suggesting that strong negative reactions and a heightened sense of crisis on X tended to temporally precede similar emotional patterns observed in news reporting. By contrast, in September, sadness and surprise showed higher transfer entropy from news to X, whereas joy was more strongly transferred from X to news.

To further examine temporal variations in directional relationships, we conducted a sliding-window transfer entropy analysis focusing on the event period (August 15–September 15) (see Figure 8). This period includes a phase where key emotional transitions, such as the crossover between fear and anticipation, were observed.

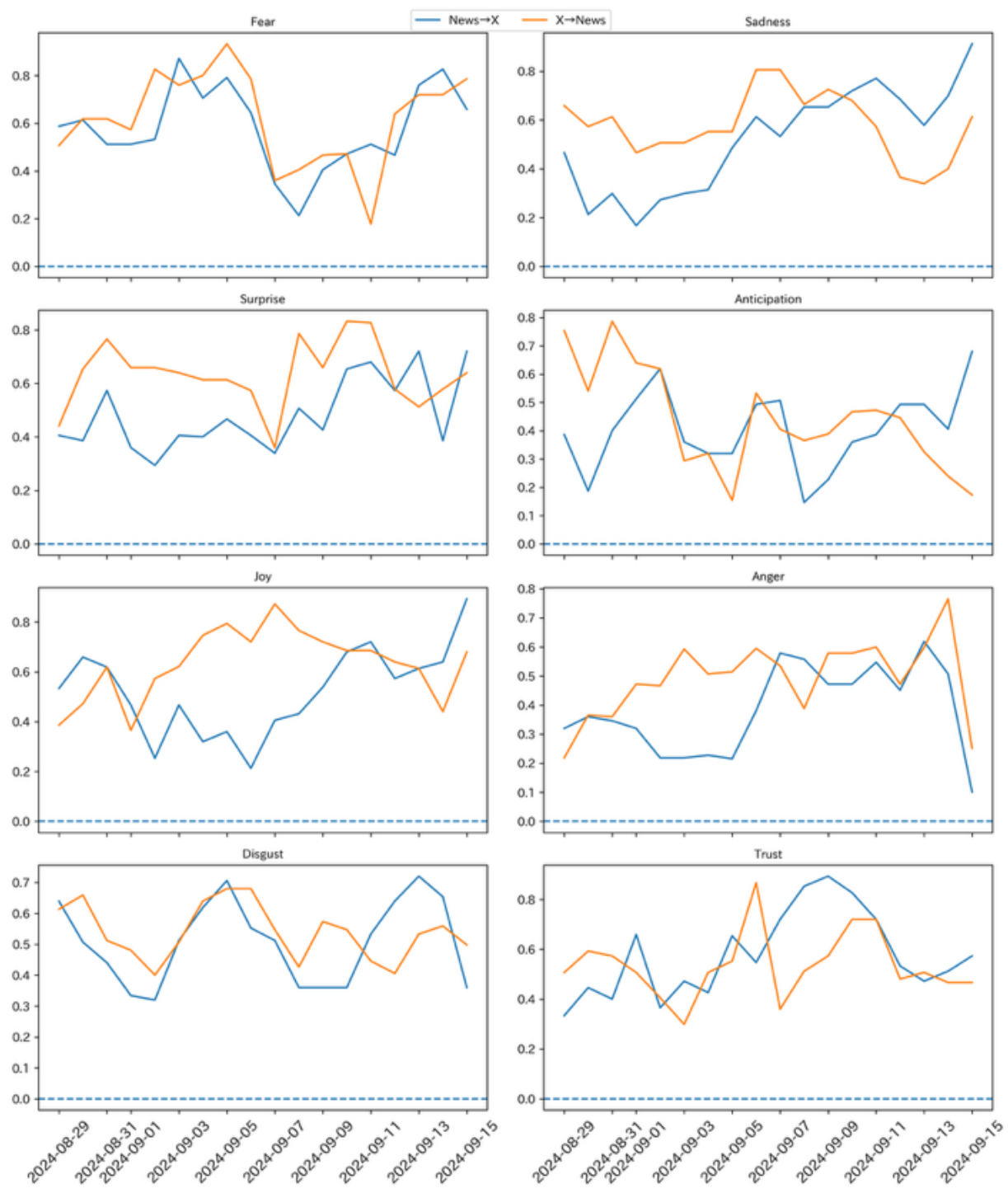

**Figure 8.** Sliding-window transfer entropy during the event period (August 15–September 15)

As shown in Figure 8, fear exhibited a pattern in which X-to-news transfer entropy was relatively higher in the early phase, while news-to-X became more prominent in the later phase, indicating a temporal shift in directional relationships. Also, sadness showed a more sustained increase in news-to-X transfer entropy during the later phase.

## Discussion

We propose a framework that integrates Plutchik's eight basic emotions with machine learning–based detection, time-series analysis, and transfer entropy to investigate temporal shifts and directional relationships of emotions across social and mass media. Its applicability is demonstrated with the 2024 "Reiwa Rice Riot" in Japan. The analysis reveals that emotional fluctuations appeared earlier on X than in news reporting, providing new insights into the temporal relationship between emotions in news and social media. These findings also demonstrate the framework's effectiveness in capturing intermedia emotional shifts and suggest its potential for broader application to other cases of social crises and public controversies.

As shown in Figure 2, a sharp increase in the number of X posts occurred from late August onward, followed shortly by a rise in the number of news articles. The number of posts on X began to surge around mid-August, featuring frequent content such as “reports of panic buying in specific regions,” “information sharing on distribution delays,” and “concerns over price hikes.” Meanwhile, the number of news articles started to increase with a delay of approximately 10 days, reaching its peak around early September. This suggests that discussions on X emerged earlier than increases in news article output, although both may have been driven by common external events. Previous studies have shown that news organizations tend to publish evidence-based content,[33] suggesting that on-the-ground voices and rumors may have been subject to some form of verification. However, the imbalance in sample sizes between social media posts and news articles may have influenced the stability of the estimated emotional trends, particularly for news data with fewer observations.

As observed in Figures 3 and 4, both platforms initially showed fear as the most prominent emotion, which was later surpassed by anticipation. As summarized in Table A1 in Appendix A, developments such as panic buying following the major earthquake advisory, governmental explanations regarding supply stabilization, and the beginning of new rice distribution may have contributed to the observed emotional shifts from fear to anticipation. Previous studies have shown that emotional expressions and topics on social media shift in response to changes in government policy,[34] and our findings align with such results. Using Plutchik’s eight basic emotions also enabled a more nuanced view than traditional positive–negative sentiment classification.

To further interpret these patterns, our results for August and September (Figures 5–7) suggest the following pattern: in August, topics concerning rice shortages were widely shared on X, and this may have been followed by an increase in news coverage. These observations suggest a phase in which social media may have played a more prominent role in the early dissemination of information. However, by September, the discourse surrounding rice shortages had matured, and news media appeared to engage in more in-depth reporting over time. Social media, compared to news media, is also known to respond more rapidly to new information,[35] and new analytical content or unexpected reports in news articles may have been associated with emotional reactions such as surprise and sadness on X. As a result, the dominant directional relationships for some emotions appeared to shift in September, with news-to-X patterns becoming more prominent. However, this interpretation should be treated with caution, as these observed emotional patterns may reflect platform-specific characteristics of X, such as its user demographics and rapid information diffusion mechanisms, rather than generalizable dynamics across all social media platforms.

Also, the data used in this study exhibit a substantial imbalance in sample size, with the number of X posts greatly exceeding that of news articles. This imbalance may affect the estimation of Transfer Entropy. To examine this issue, we conducted a supplementary analysis in which, for each day, an equal number of X posts were randomly sampled to match the number of news articles (equal-size sampling), and this procedure was repeated 500 times. The results showed that, although the absolute values of Transfer Entropy exhibited some variation, the overall directional patterns were largely preserved (see Appendix E). Therefore, while the sample size imbalance may influence the magnitude of the estimates, it does not fundamentally alter the main finding of this study, namely the tendency for emotional changes on X to temporally precede similar emotional patterns in news media under certain conditions.

In addition, as shown in Figure 8, the sliding-window transfer entropy analysis revealed that directional relationships fluctuated dynamically over time, with the dominant direction alternating between X-to-news and news-to-X for several emotions. The sliding-window analysis therefore offers a more nuanced understanding of how directional relationships evolve over time, particularly in rapidly developing social events.

The proposed framework not only captured the temporal shifts of emotions but also visualized how temporal relationships between social and news media evolved. These insights, which could not have been obtained through simple sentiment classification, highlight the framework's value for systematically analyzing intermedia emotional relationships in social events. However, the generalizability of the observed emotional patterns should be considered with caution. While the analytical framework proposed in this study may be applicable to other social crises and public controversies, the observed shift from fear to anticipation may have been influenced by the unique characteristics of the "Reiwa Rice Riot" and the Japanese sociocultural context.

This study has several limitations. The analysis was limited to X, although during social crises or controversies, discussions are also actively conducted on other platforms such as Instagram and Facebook.[36-37] Analyses confined to a single platform inevitably introduce observational bias. Therefore, without multi-platform analysis, it is difficult to capture the full picture of emotional diffusion. It should also be noted that the emotion classification model was trained on social media text (WRIME dataset), and its performance on news articles with different linguistic characteristics may be affected by domain differences.

Additionally, although transfer entropy is a useful metric for quantifying directional emotional flow between time series, it does not establish true causality. This is because external factors, such as official government announcements or breaking news on TV and newspapers, may simultaneously influence both social media and news media. Therefore, the observed directional relationships may be affected by such confounding variables and should be interpreted with caution. Furthermore, the stationarity tests (ADF and KPSS) indicated that while most news-related emotional series were stationary, X-based emotional series exhibited non-stationarity and structural breaks, reflecting their event-driven nature. Although emotional time series on X exhibit stationarity over longer and more stable periods (see Appendix F), the data analyzed in this study were collected during a specific event period, which is characterized by rapid changes and external shocks. Additionally, even within each sliding window, strict stationarity cannot be fully guaranteed. Therefore, the results should be interpreted as reflecting temporal dependencies and evolving information flow rather than strict causal relationships.

Future work should include cross-platform data collection and analysis to compare how emotional dynamics differ across social media platforms. Moreover, it could also explore cross-emotional transitions (e.g., disgust preceding anger), providing a more comprehensive understanding of emotional dynamics. It is also necessary to model the sequential process of the present case using causal inference methods and quantitatively examine the extent to which emotional trends on social media contribute to policy decisions or supply chain responses. In addition, applying the proposed framework to diverse social and cultural contexts and different types of events would be important to evaluate its generalizability.

The main contribution of this study is to show that shifts in emotions on X tended to temporally precede similar emotional patterns observed in news media. Our findings suggest that patterns in emotional expressions on X provide insights into temporal shifts in emotions during social crises and the temporal precedence of emotional changes across social and news media.

**Data availability**

The datasets analyzed during this study are available from the corresponding author upon reasonable request.

## Appendix

Appendix A. Timeline of major developments during the “Reiwa Rice Riot”

Table A1. Major developments during the “Reiwa Rice Riot”

| Period | Major event |
|---|---|
| Late July 2024 | Reports concerning declining private rice inventories and rising rice prices increased, leading to growing nationwide concerns over supply shortages. |
| August 8, 2024 | Following the announcement of the major earthquake advisory, panic buying and purchase restrictions expanded due to increased demand for emergency stockpiling. |
| August 27, 2024 | In addition to the beginning of new rice distribution, reports emerged regarding governmental measures and explanations by the Ministry of Agriculture, Forestry and Fisheries aimed at stabilizing supply and normalizing distribution. |
| Mid-September 2024 | Reports concerning the expansion of new rice distribution and prospects for supply recovery increased, and improvements in store inventories were also observed. |

Appendix B: Emotion classification model performance on X

The performance of the emotion classification model was evaluated. The results are as follows:

Accuracy: 0.7476

F1-score: 0.74399

Precision: 0.75095

Recall: 0.74757

Appendix C. Emotion classification model performance on news articles

To evaluate the performance of the BERT classifier on news articles, the following procedure was conducted. Two annotators labeled 100 articles on an ordinal four-point scale for eight emotions: joy, sadness, anticipation, surprise, anger, fear, disgust, and trust. The agreement rate between the dominant emotions identified by human annotations and BERT predictions was 53%.

## Appendix D. Robustness to discretization (bin sensitivity analysis)

To examine the sensitivity of the results to the choice of discretization, we varied the number of bins used in the discretization step. Specifically, Transfer Entropy was recalculated using different bin settings (3, 4, and 5 bins) for the combined data from August and September 2024.

For each bin setting, we compared the direction of information flow (News→X vs. X→News) for each emotion. Table D1 summarizes the number of conditions under which each direction was dominant.

Table D1. Directional dominance of Transfer Entropy across different bin settings (3, 4, and 5 bins). Values indicate the number of conditions in which each direction (News→X or X→News) was dominant for each emotion.

| Emotion | News→X dominant | X→News dominant |
|---|---|---|
| Fear | 0 | 3 |
| Sadness | 1 | 2 |
| Surprise | 1 | 2 |
| Anticipation | 0 | 3 |
| Joy | 0 | 3 |
| Anger | 1 | 2 |
| Disgust | 0 | 3 |
| Trust | 0 | 3 |

## Appendix E. Robustness check using equal-size sampling

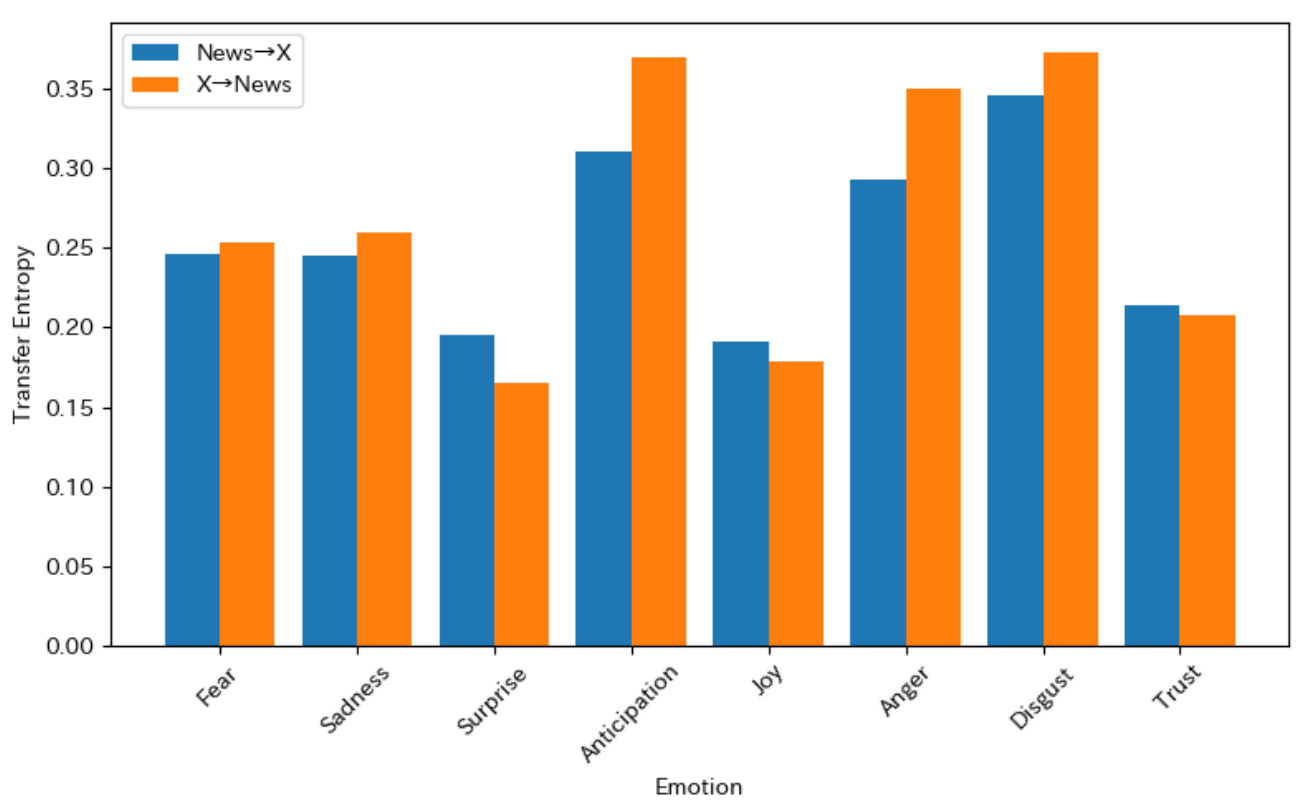

Figure E1. Transfer Entropy results based on day-matched equal-size sampling (500 iterations). Bars represent the mean values across iterations for each emotion.

## Appendix F. Supplementary analysis on stationarity using Twitter data

To examine whether the non-stationarity observed in the main analysis is specific to the event period, we conducted an additional analysis using baseline Twitter data unrelated to the event. Specifically, we collected a 1% random sample of all Twitter posts from January 1 to June 30, 2022, and randomly selected 1,000 posts per day. These data were processed using the same preprocessing and emotion estimation procedures as in the main analysis, based on Plutchik's eight basic emotions. We then constructed time series by calculating the daily mean scores for each emotion and evaluated stationarity using the Augmented Dickey–Fuller (ADF) and KPSS tests. As a result, many of the emotional time series were found to be stationary. These findings suggest that emotional time series during non-event periods are generally stationary.

Table F1. Results of stationarity tests for emotional time series based on baseline Twitter data (January–June 2022).

| Emotion | ADF p-value | KPSS p-value |
|---|---|---|
| Fear | $p < .01$ | 0.100 |
| Sadness | $p < .01$ | 0.100 |
| Surprise | $p < .01$ | 0.025 |
| Anticipation | $p < .01$ | 0.070 |
| Joy | $p < .01$ | 0.100 |
| Anger | 0.027 | 0.010 |
| Disgust | $p < .01$ | 0.010 |
| Trust | $p < .01$ | 0.025 |

## Author contributions

E.M. conceived the experiment in consultation with M.C. and F.T. E.M. conducted the experiment. E.M. analyzed the results in consultation with M.C. and F.T. F.T. provided the data. All authors reviewed the manuscript.

## Declarations

### Competing interests

The authors declare no competing interests.

### Funding

This research was supported by JST ERATO (Grant Number JPMJER2502).